\documentclass[a4paper,11pt]{article}
\usepackage{amsmath}
\usepackage{bm}	
\usepackage{tikz-feynman}
\usepackage{tikz}
\usepackage{float}
\usepackage{caption}
\usepackage{subcaption}
\usepackage{pdflscape}
\usepackage{upgreek}

\usepackage[pdftex]{hyperref}
\hypersetup{
  bookmarksopen=true, breaklinks=true, debug=true, %
  colorlinks=true, linkcolor=blue, citecolor=blue, urlcolor=blue
}

\newcommand\had{\mathrm{h}}
\newcommand\lep{\ell}

\newcommand\nukl{\mathrm{N}}
\newcommand\fin{\mathrm{X}}

\def\phih{\phi_\had}

\def\PhT{P_\text{T}}
\def\khT{k_\text{T}}
\newcommand{\A}[2]{A_{\rm #1}^{#2}}
\def\acos{A_\text{UU}^{\cos\phi_\had}}
\def\accos{A_\text{UU}^{\cos2\phi_\had}}

\def\afx{A_\text{XU}^{f(\phi_\had)}}

\def\GeVc{~\text{GeV}/c}

\newcommand{\dif}[1]{{\rm d} #1 \,}
\newcommand{\eps}{\varepsilon}

\title{Transverse momentum and azimuthal angle distributions of hadrons in DIS at COMPASS}

\author{Vendula Bene\v{s}ov\'{a}$^{a,*}$ on behalf of the COMPASS Collaboration}

\def\email{$^{*}$\href{mailto:vendula.benesova@cern.ch}{vendula.benesova@cern.ch}}

\def\affiliation[a]{$^{a}$Charles University, Prague, Czech Republic}

\def\abstract{The COMPASS experiment plays an important role in studies of nucleon structure. Measurements performed in 2016–2017 with an unpolarised liquid hydrogen target and a longitudinally polarised 160 GeV/$c$ muon beam provide a high-statistics data sample for the analysis of semi-inclusive deep-inelastic scattering, giving access to transverse-momentum-dependent parton distribution and fragmentation functions.

The 2016 data set was analysed to determine the multiplicities of hadrons produced in DIS differential in $x$, $Q^2$ and $\PhT$, as well as the amplitudes of azimuthal modulations in the hadron azimuthal-angle distributions. In particular, the unpolarised asymmetries $A^{\cos \phi_\text{h}}_\text{UU}$ (related to the Cahn effect), $A^{\cos 2\phi_\text{h}}_\text{UU}$ (sensitive to the Boer--Mulders function), and the beam-spin asymmetry $A^{\sin \phi_\text{h}}_\text{LU}$ were extracted in 4-dimensional bins matching those of the multiplicities, and also integrated over some of the variables.}

\def\FullConference{Presented on 5th May 2026 at 33rd International Workshop on Deep Inelastic Scattering (DIS2026) in Bologna, Italy.}

\makeatletter
    \def\@maketitle{%
  \newpage
  \null
  \vskip 2em%
  \begin{center}%
  \let \footnote \thanks
    {\LARGE \@title \par}%
    \vskip 1.5em%
    {\large
      \lineskip .5em%
      \begin{tabular}[t]{c}%
        \@author
      \end{tabular}\par}%
      \vskip .5em%
    {\large
      \lineskip .5em%
      \begin{tabular}[t]{c}%
        \affiliation[a]
      \end{tabular}\par}%
      \vskip .5em%
    {\large
      \lineskip .5em%
      \begin{tabular}[t]{c}%
        \email
      \end{tabular}\par}%
      \vskip 1em%
    {\small
      \lineskip .5em%
      \begin{tabular}[t]{c}%
      \parbox{0.8\textwidth}{\centering
        \FullConference}
      \end{tabular}\par}%
      \vskip 1em%
      \subsection*{Abstract}
    {\large
      \lineskip .5em%
      \parbox{0.8\textwidth}{
       \begin{flushleft}
      \begin{abstract}%
      \end{abstract}
      \end{flushleft}\par}}%
  \end{center}%
  \par
  \vskip 2em}
\makeatother

\begin{document}
\maketitle

\section{Introduction}

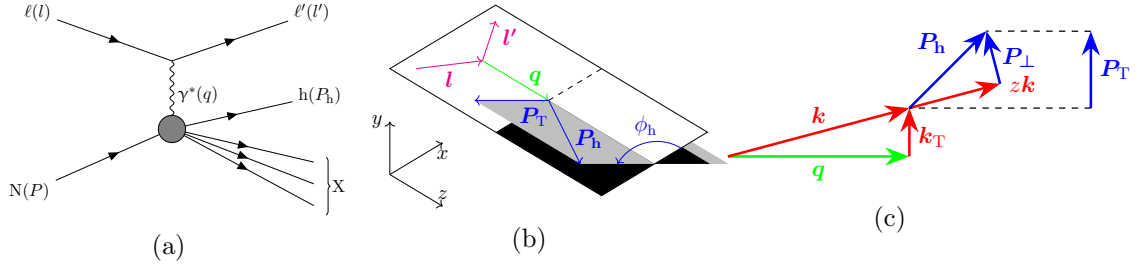
\begin{figure}[tbp]
    \begin{subfigure}[tbp]{0.32\textwidth}
    \scalebox{0.6}{
\begin{tikzpicture}
  \begin{feynman}
    \vertex (i1) {\( \lep(l) \)};
    \vertex [right=1.9cm of i1] (i2);
    \vertex [below right=of i2] (i3);
    \vertex [above right=of i3] (i4);
    \vertex [right=1.5cm of i4] (i5) {\( \lep'(l') \)};
    \vertex [below =of i3] (o1);
    \vertex [left=of o1] (o2);
    \vertex [right=of o1] (o3);
    \vertex [below left=of o2] (b) {\( \nukl(P) \)};
    \vertex [below right=2.3cm of o3] (a) {\( \)};
    \vertex [above=0.5cm of a] (a1) {\( \)};
    \vertex [above=0.5cm of a1] (a2) {\( \)};
    \vertex [above=1.5cm of a2] (a3) {\( \had(P_\had) \)};
    
    \diagram* {
      (i1) -- [fermion] (i3) -- [fermion] (i5),
      (i3) -- [boson, edge label=\( \upgamma^*(q) \)] (o1),
      (b) -- [fermion] (o1),
      (o1) -- [fermion] (a),
      (o1) -- [fermion] (a1),
      (o1) -- [fermion] (a2),
      (o1) -- [fermion] (a3),
    };
    \draw[fill=gray] (o1) circle [radius=0.3cm];
    \draw [decoration={brace}, decorate] (a2.north east) -- (a.south east)
          node [pos=0.5, right] {\(\fin\)};
  \end{feynman}
\end{tikzpicture}}
\caption{~}\label{fig:SIDIS}
\end{subfigure}
\begin{subfigure}[tbp]{0.32\textwidth}
    \scalebox{0.7}{
 \begin{tikzpicture}
  \begin{feynman}

    \coordinate (AA) at (2,-1.2);
    \coordinate (BB) at (4,-2.4);
    \coordinate (CC) at ($(BB) + (1,0.6)$);
    \coordinate (X) at ($(AA) + (2,1.2)$);
    \coordinate (DD) at ($(AA) + (1,0.6)$);
    \vertex (c1) at ($(DD)$);
    \vertex (c2) at ($(DD) - (1.25,-0.75)$);
    \coordinate (c3) at ($(DD) - (1.75,-1.05)+ (0.75,0.45)$);
    \coordinate (c4) at ($(DD) - (1.75,-1.05)- (0.75,0.45)$);
    \coordinate (EE) at ($(AA) + (0.4,-0.6) + (2,1.2)$);
    \coordinate (FF) at ($(BB) + (0.4,-0.6) + (2,1.2)$);
    \coordinate (EEE) at ($(AA) + (0.3,-0.45) + (1.75,1.05)$);
    \coordinate (FFF) at ($(BB) + (0.3,-0.45) + (1.75,1.05)$);
    \coordinate (GG) at ($(BB) + (1,0.6)$);
    \coordinate (HH) at ($(AA) + (1,0.6)$);
    \coordinate (A) at (0,0);
    \coordinate (B) at (4,-2.4);
    \coordinate (C) at ($(B) + (2,1.2)$);
    \vertex (az1) at ($(B) + (1.5,0.9)$);
    \coordinate (D) at ($(A) + (2,1.2)$);
    \fill[lightgray] (EE) -- (FF) -- (GG) -- (HH) -- cycle;    
    \fill[black] (EEE) -- (FFF) -- (GG) -- (HH) -- cycle;
    \draw[fill=white] (A) -- (B) -- (C) -- (D) -- cycle;
    \fill[black] (AA) -- (BB) -- (CC) -- (DD) -- cycle;

    \coordinate (E) at ($(AA) + (-0.4,0.6)$);
    \coordinate (F) at ($(BB) + (-0.4,0.6)$);
    \coordinate (G) at ($(BB) + (1,0.6)$);
    \coordinate (H) at ($(AA) + (1,0.6)$);
    \vertex (az2) at ($(BB) + (0.3,0.6)$);
    
    \fill[lightgray] (E) -- (F) -- (G) -- (H) -- cycle;
  
    \draw[->, line width=0.5pt] (0,-2) -- (1,-1.4) node[below]{$x$};
    \draw[->, line width=0.5pt] (0,-2) -- (0,-1) node[left]{$y$};
    \draw[->, line width=0.5pt] (0,-2) -- (1,-2.6) node[above]{$z$};
    \draw[->, line width=0.5pt, color=magenta] (c2) -- (c3) node[below right,font=\large]{$\boldsymbol{l'}$};
    \draw[->, line width=0.5pt, color=magenta] (c4) -- node[below,font=\large]{$\boldsymbol{l}$} (c2);
    \draw[->, line width=0.5pt, color=green] (c2) -- (c1) node[pos=0.8, above,font=\large]{$\boldsymbol{q}$}; 
    \draw[->, line width=0.5pt, color=blue] (c1) -- node[pos=0.6, right,font=\large]{$\boldsymbol{P_\had}$} (F); 
    \draw[->, line width=0.5pt, color=blue] (c1) -- node[pos=0.2, below,font=\large]{$\boldsymbol{\PhT}$} (E);
    \draw [->, bend right=45, color=blue]  (az1) to node[above,font=\large]{$\phih$} (az2);
    \draw [dashed, line width=.15pt] (DD) -- (X);
    
  \end{feynman}
\end{tikzpicture}}
\caption{~}\label{fig:GNS}
\end{subfigure}
\begin{subfigure}[tbp]{0.32\textwidth}
    \scalebox{0.8}{
        \begin{tikzpicture}
  \begin{feynman}
    \vertex (i2);
    \vertex [right=3cm of i2] (i1);
    \vertex [above=0.8cm of i1] (a);
    \vertex [right=of a] (c);
    \vertex [above=0.40cm of c] (c1);
    \vertex [above right=1.8cm of a] (c2);

    \vertex [right=of c] (d);
    \vertex [above=1.27cm of d] (d1);

    \draw [-{Stealth[length=4mm, width=3mm]}, line width=1.2pt, color=green] (i2) -- node [below] {\( \boldsymbol{q} \)} (i1);
    \draw [-{Stealth[length=4mm, width=3mm]}, line width=1.2pt, color=red] (i2) -- node [above] {\( \boldsymbol{k}\)} (a);
    \draw [-{Stealth[length=4mm, width=3mm]}, line width=1.2pt, color=red] (i1) -- node [pos=0.8, below right] {\( \boldsymbol{\khT}\)} (a);
    \draw [-{Stealth[length=4mm, width=3mm]}, line width=1.2pt, color=red] (a) -- (c1)   node [right] {\( z\boldsymbol{k}\)};
    \draw [-{Stealth[length=4mm, width=3mm]}, line width=1.2pt, color=blue] (a) --  node [pos=0.6, above left] {\( \boldsymbol{P_\had}\)} (c2);
    
    \draw [-{Stealth[length=4mm, width=3mm]}, line width=1.2pt, color=blue] (c1) -- node [right] {\( \boldsymbol{P_\perp}\)} (c2);
    \draw [-{Stealth[length=4mm, width=3mm]}, line width=1.2pt, color=blue] (d) -- node [right] {\( \boldsymbol{\PhT}\)} (d1);

    \draw [dashed, line width=.6pt] (d1) -- (c2);
    \draw [dashed, line width=.6pt] (d) -- (a);
  \end{feynman}
\end{tikzpicture}}
\caption{~}\label{fig:transvers_momenta}
\end{subfigure}

    \caption{\textbf{(a)} Feynman diagram of the SIDIS process at the tree level. \textbf{(b)} Definition of the transverse momentum $\PhT$ and azimuthal angle $\phih$ of the final state hadron in the $\upgamma^*$NS. \textbf{(c)} Origin of $\boldsymbol{\PhT}$ in SIDIS process from the intrinsic transverse momentum of quark $\khT$ inside the nucleon and from the momentum gained in the fragmentation process $P_\perp$. }
    \label{fig:GNSSIDIS}
\end{figure}

Hadrons produced in the current fragmentation region in semi-inclusive DIS ($\lep\nukl\rightarrow\lep'\had\fin$ in Fig.~\ref{fig:SIDIS}) carry information about the nucleon structure, which can be accessed through their transverse momentum $\PhT$ and azimuthal angle $\phih$, defined in the virtual photon--nucleon system ($\upgamma^*$NS) as shown in Fig.~\ref{fig:GNS}. Non-zero transverse momentum originates from the intrinsic transverse momentum of quark $\khT$ inside the nucleon and from the momentum gained in the fragmentation process $P_\perp$ as illustrated in Fig.~\ref{fig:transvers_momenta}.

The azimuthal angle dependent cross-section of the production of a hadron $\had$ in deep inelastic scattering of a lepton off an unpolarised target in the one photon exchange approximation is~\cite{Bacchetta:2006tn}
\begin{equation}\label{eq:xsec}
    \frac{\dif{\sigma^\had}}{\dif{\PhT^2} \dif{x} \dif{y} \dif{z} \dif{\phih}} = \sigma_0\left(1
        +\eps_1  \A{UU}{\cos\phih}\cos\phih
        + \eps_2 \A{UU}{\cos2\phih}\cos2\phih
        + \lambda \eps_3 \A{LU}{\sin\phih}\sin\phih \right),
\end{equation}
where $x$ is the Bjorken variable, $y$ is the fraction of the beam energy carried by the virtual photon, $z$ is the fraction of the photon energy transferred to the hadron, and $\phih$ is the azimuthal angle of the hadron in GNS. The beam polarisation is indicated with $\lambda$ and the terms $\eps_i$ are kinematic factors depending on $y$, defined as:
\begin{equation}\label{eq:epsilons}
	\eps_1 = \frac{2(2-y)\sqrt{(1-y)}}{1+(1-y)^2}, \hspace{1.5cm}
    \eps_2 = \frac{2(1-y)}{1+(1-y)^2}, \hspace{1.5cm}
    \eps_3 = \frac{2y\sqrt{1-y}}{1+(1-y)^2}.
\end{equation}
 The observables obtained by fitting the cross-section on measured $\phih$-distributions are amplitudes $A_\text{XU}^{f( \phih)}$ -- \textit{azimuthal asymmetries} -- defined as ratios of the structure functions $F_\text{XU}^{f( \phih)}$:
\begin{equation}
     A_\text{XU}^{f( \phih)}\bigl(x,z,\PhT^2,Q^2\bigr)\equiv\frac{F_\text{XU}^{f( \phih)}}{F_\text{UU}}\ , \qquad F_\text{UU}= F_\text{UU,T}+\varepsilon F_\text{UU,L}\ .
\end{equation}

Only $F_\text{UU}$ contributes to the cross-section integrated over $\phih$, which describes $\PhT^2$-distributions~\cite{Anselmino:2013lza}:
\begin{equation}\label{eq:xsecint}
     \frac{\dif{\sigma^\had}}{\dif{x}\dif{Q^2} \dif{z} \dif{\PhT^2}} 
     = \frac{2\pi^2\alpha^2}{xyQ^2} \frac{\left[1+(1-y)^2\right]}{y^2}F_\text{UU}\ .
\end{equation}

Cleaner observables are the $\PhT^2$-dependent multiplicities, obtained by normalising the cross section in Eq.~\ref{eq:xsecint} to the inclusive DIS cross-section. Experimentally, the measurement is performed by counting the number of produced hadrons, $N^{\had^\pm}$, and DIS events, $N^\text{DIS}$, and normalising to the bin widths of the hadronic variables $\Delta\PhT^2$ and $\Delta z$ as follows:

\begin{equation}
     M^{\had^{\pm}}(x, Q^2, z, \PhT^2) \equiv \frac{\dif{\sigma^\had}}{\dif{ x}\dif{Q^2} \dif{ z} \dif{ \PhT^2}} \Big/ \frac{\dif{\sigma^\text{DIS}}}{\dif{x} \dif{Q^2}}
    = 
    \frac{1}{\Delta \PhT^2 \Delta z}
    \frac{N^{\had^{\pm}} (x, Q^2, z, \PhT^2)}
    {N^\text{DIS} (x, Q^2)}\ .
\end{equation}

Structure functions can be interpreted as weighted convolutions of TMD-PDFs and TMD-FFs.

\section{Data sample and COMPASS spectrometer setup}

The COMPASS experiment is a fixed-target experiment located at the M2 beamline of the CERN Super Proton Synchrotron (SPS). Data were collected between 2002 and 2022 as part of a broad physics programme aimed at studying the structure and spectroscopy of hadrons with high-intensity muon and hadron beams. The analysis of unpolarised SIDIS presented in this work is based on data collected in 2016 using a liquid hydrogen target and a longitudinally polarised $\upmu^\pm$ beam with a momentum of 160 \GeVc. A similar analysis of the spin-independent part of the SIDIS cross section was previously performed using data collected in 2004 and 2006 with a polarised $^6$LiD target \cite{Adolph:2014pwc,Agarwala_2020,Adolph_2013}.

\section{Key Monte-Carlo based corrections}\label{sec:corrections}

Unlike analyses of the polarised part of the SIDIS cross section, studies of the unpolarised SIDIS cross section rely heavily on Monte Carlo corrections. In addition to acceptance corrections, which account for inhomogeneous detector efficiency, finite geometrical coverage, and the finite resolution of the spectrometer, further corrections are required to avoid biases arising from contamination by exclusive processes and radiative effects.

In the case of the treatment of the exclusive background, the majority of hadrons originating from exclusive processes are suppressed by applying a cut on the exclusive peak, namely by requiring $z_\text{tot}<0.9$ for events with 2 and only 2 reconstructed hadrons with opposite charge detected in the spectrometer. The rest of the correction is carried out by a HEPGEN-generator-based MC sample \cite{Charchula:1994kf}. After reconstruction of the MC-sample, the normalised $\phih$ distributions in HEPGEN are directly subtracted from the measured ones.

Radiative effects are corrected for using DJANGOH-generator-based MC samples \cite{sandacz2012hepgengeneratorhard}. Two MC samples are produced: one including radiative effects and one generated at the Born level, without radiative effects. The radiative correction factors are determined as the ratio of the hadron yields in the two samples, each normalized to the corresponding number of DIS events. The measured $\phih$ distributions and multiplicities are then corrected by dividing them by these radiative correction factors.

More details on the correction procedures and their effects on the results can be found in Ref.~\cite{Benesova:2024cfk,Matousek:2026ebs,COMPASS:2024gje}. The aforementioned effects are significantly reduced in polarised measurements, as they largely cancel in subtractions of cross-sections corresponding to different target polarisation configurations.

The correction procedure for previously published unpolarised SIDIS measurements on isoscalar targets is not uniform across different analyses. For azimuthal asymmetries, the contribution from exclusive background was subtracted ad-hoc after publishing the first not corrected results at the asymmetry level using HEPGEN generator-based MC simulations, without applying a cut on the exclusive peak. Dependencies of radiative effects on hadronic variables were not corrected, as initial studies based on RADGEN indicated only a small impact on the measured asymmetries \cite{Adolph:2014pwc,Agarwala_2020}.

In contrast, for $\PhT^2$-dependent multiplicities from 2004 data, no correction was applied for contamination from exclusive processes~\cite{Adolph_2013}. The analysis of 2006 data included a first HEPGEN MC-based correction; however, since the correction is based on cross-sections and without applying the cut on the exclusive peak, the correction was large and carried a large systematic error~\cite{COMPASS:2017mvk}. In both papers, radiative effects were accounted for using TERAD tables; however, these provide an inclusive correction and therefore accurately reproduce only the DIS normalisation, without describing hadron-dependent kinematic effects.

\section{Results on proton target}

The multidimensional dependences of $\acos$ are presented in Fig.~\ref{fig:4dres_cos}. The systematic uncertainty, which is represented in the plots by blue systematic bands, consists of contributions from all corrections, compatibility of different parts of the data sample, and compatibility of the results with vertexing in different parts of the target. Results of the corresponding one-dimensional analysis can be found in Refs.~\cite{Benesova:2024cfk,Matousek:2026ebs}; the only significant update is the use of modified generator settings for the DJANGOH-based MC simulation and refined systematic uncertainty.

The $\PhT^2$-dependent multiplicities, together with the impact of the radiative corrections, are shown in Fig.~\ref{fig:4dmult}. Their systematic uncertainty was not yet fully evaluated; however, the same contributions will be present as for $\afx$, and the current analysis suggests that the systematic error will not exceed 10\% and that the largest of the contributions is from the acceptance correction. Previous studies have demonstrated that the slope of the exponential fit is not significantly modified by the approximately linear behaviour of the radiative corrections \cite{Benesova:2025}. Nevertheless, the corrections are sizable in absolute value, reaching up to 40\% as visible in Fig.~\ref{fig:4dRC}.

\section{Comparison with the results on isoscalar target}

\begin{figure}[H]
        \centering
        \vspace{15pt}
        \includegraphics[width=.42\textwidth]{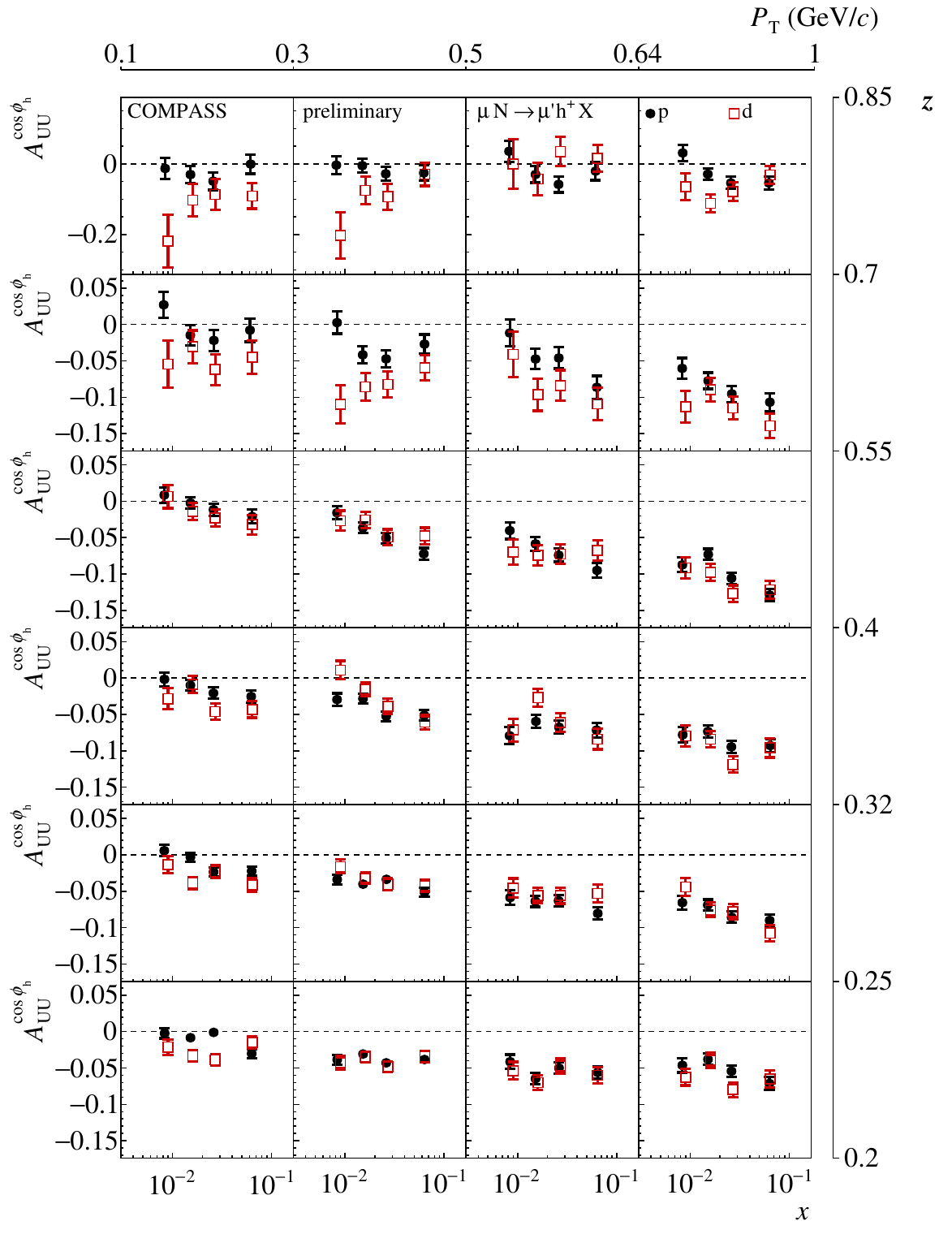}
        \includegraphics[width=0.42\textwidth]{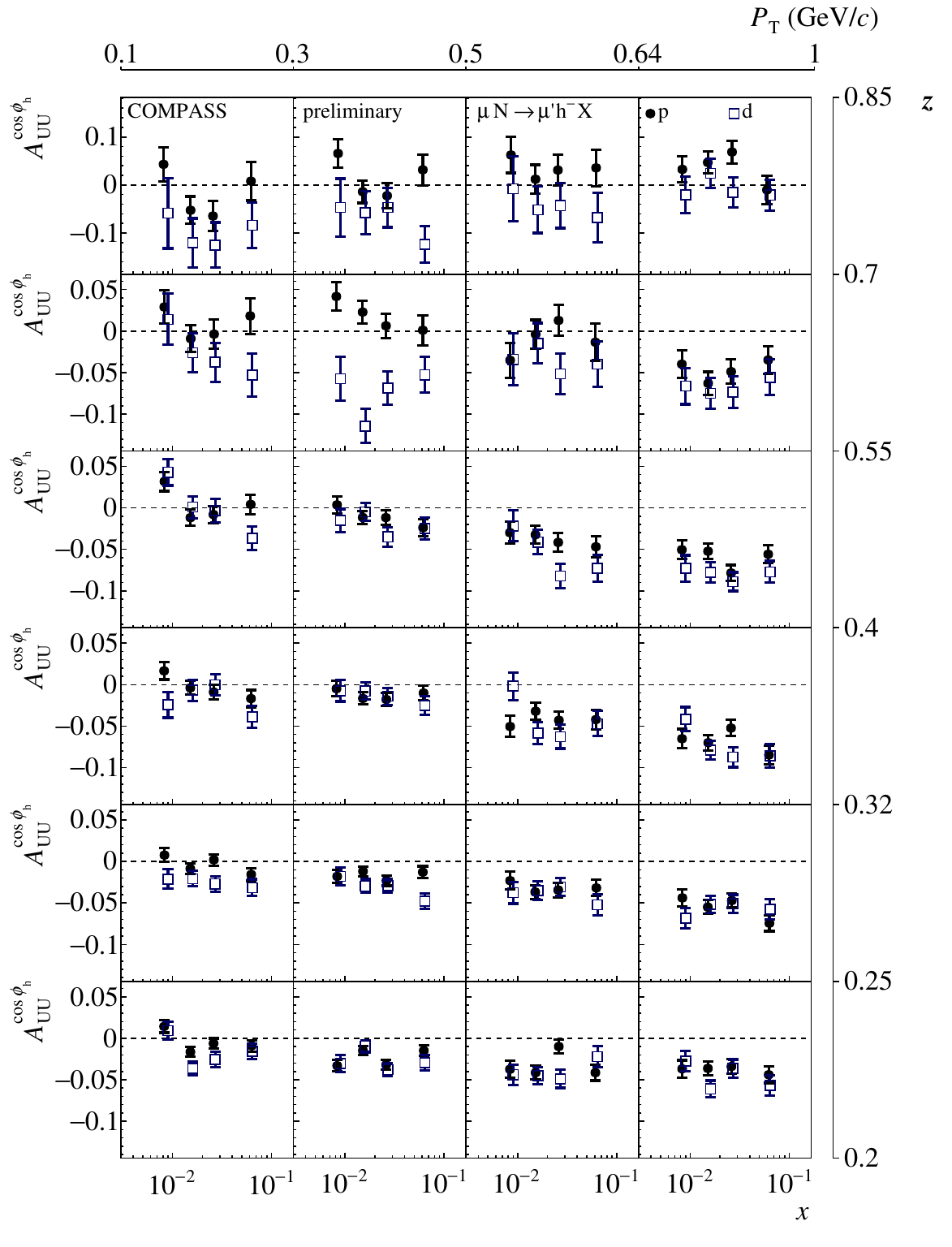}\\
        \vspace{15pt}
        \includegraphics[width=.42\textwidth]{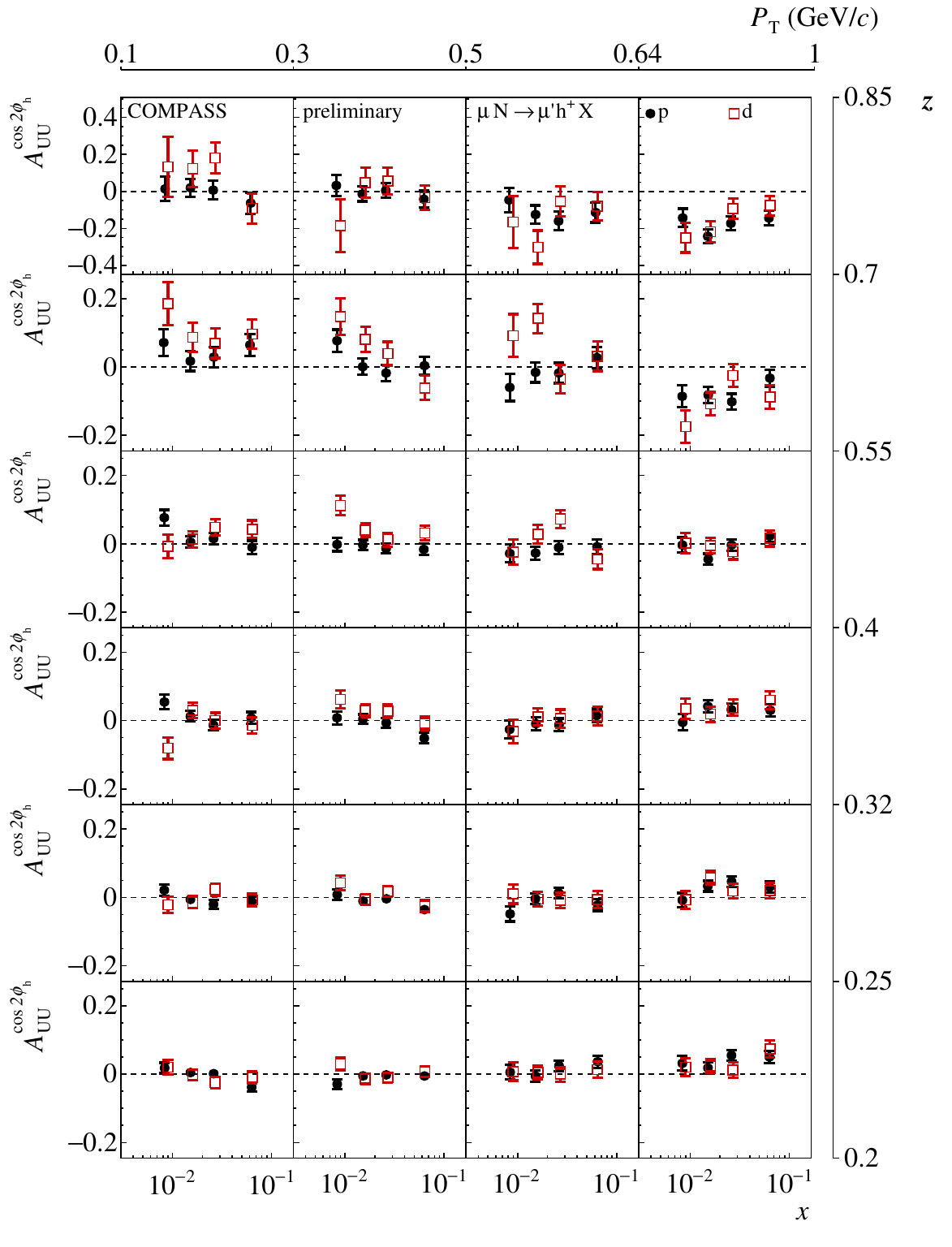}
        \includegraphics[width=0.42\textwidth]{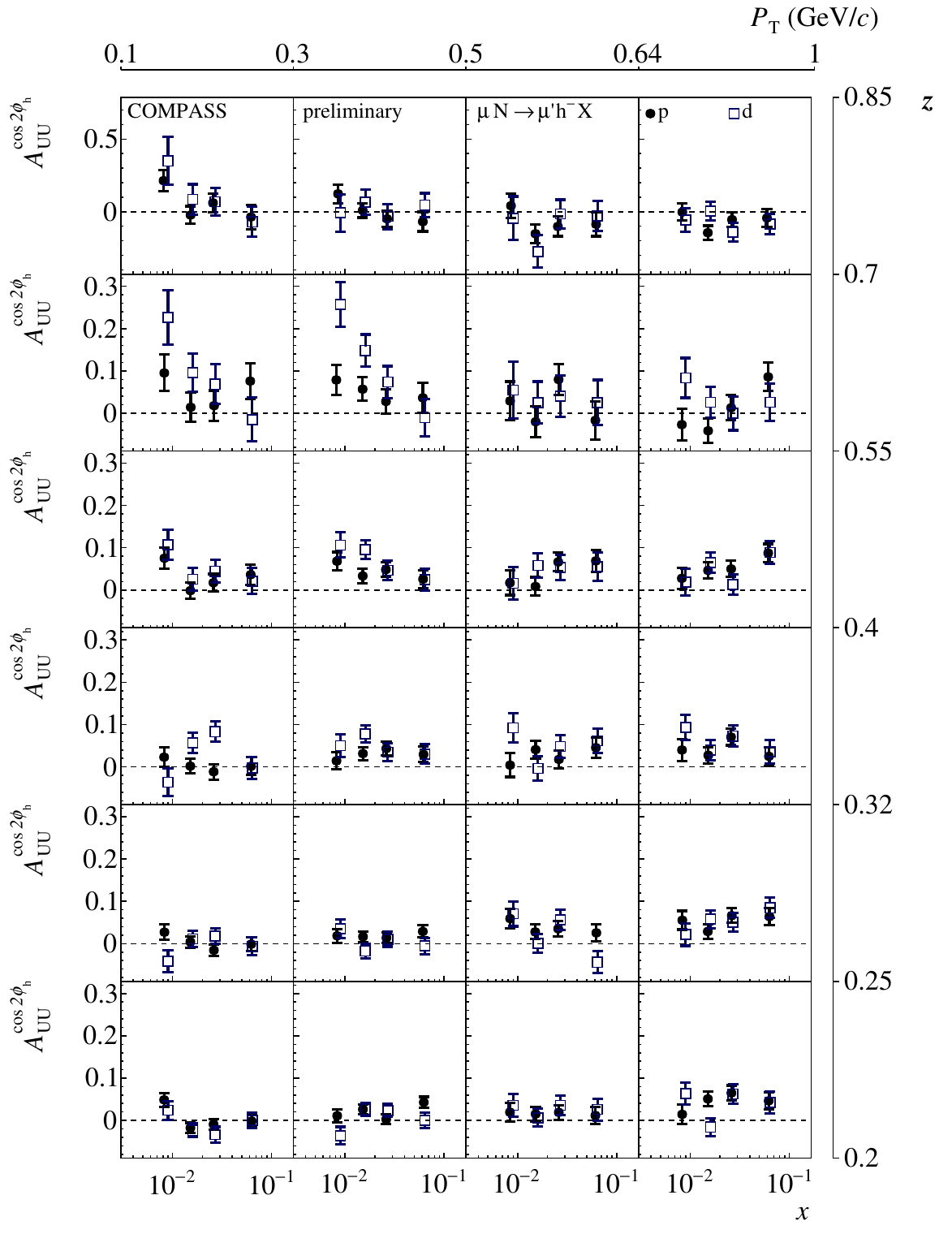}
        \vspace{15pt}
        \caption{The comparison of the COMPASS results for azimuthal asymmetries measured on proton (full dots) and deuteron (colored empty squares). Results for $\acos$ (top), for $\accos$ (bottom), positive hadrons (red), negative hadrons (blue).}\label{fig:compdeuteron}
    \end{figure}

Any difference in comparison with the results on the isoscalar target hints a flavour dependence of TMDs since we compare the up quark dominated results on proton from this work with the results on the isoscalar target (both not corrected on RC). This comparison is for $\afx$ in Fig.~\ref{fig:compdeuteron}. In theory, RE should not be target (isospin) dependent; thus, it is possible to make conclusions even when the results miss the RC. However, as described in section \ref{sec:corrections}, the correction on hadrons from decays of DVMs is done in a different manner and in case of analysis of 2004 data carries large systematics.

Care must be taken when comparing the presented $\PhT^2$-dependent multiplicities with previously published results on the deuteron target, owing to differences in the applied Monte Carlo corrections discussed in Sec.~\ref{sec:corrections}. Efforts are currently underway within the COMPASS collaboration to unite the correction procedures and ensure consistency between the data sets in future analyses.

\section{Acknowledgements}
The work of the author is supported by COST (European Cooperation in Science and Technology) Action CA24159 'SHARP'.

\begin{landscape}

\begin{minipage}{0.6\textwidth}
    \centering
    \includegraphics[width=.84\textwidth]{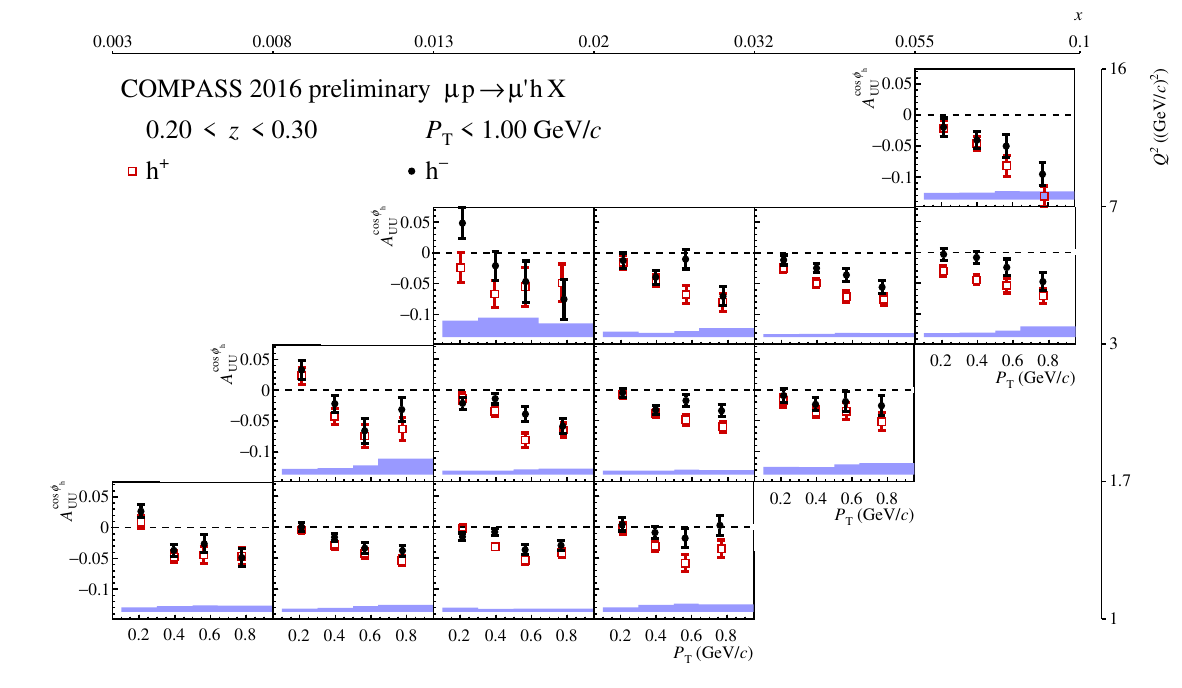}\\
    \includegraphics[width=.84\textwidth]{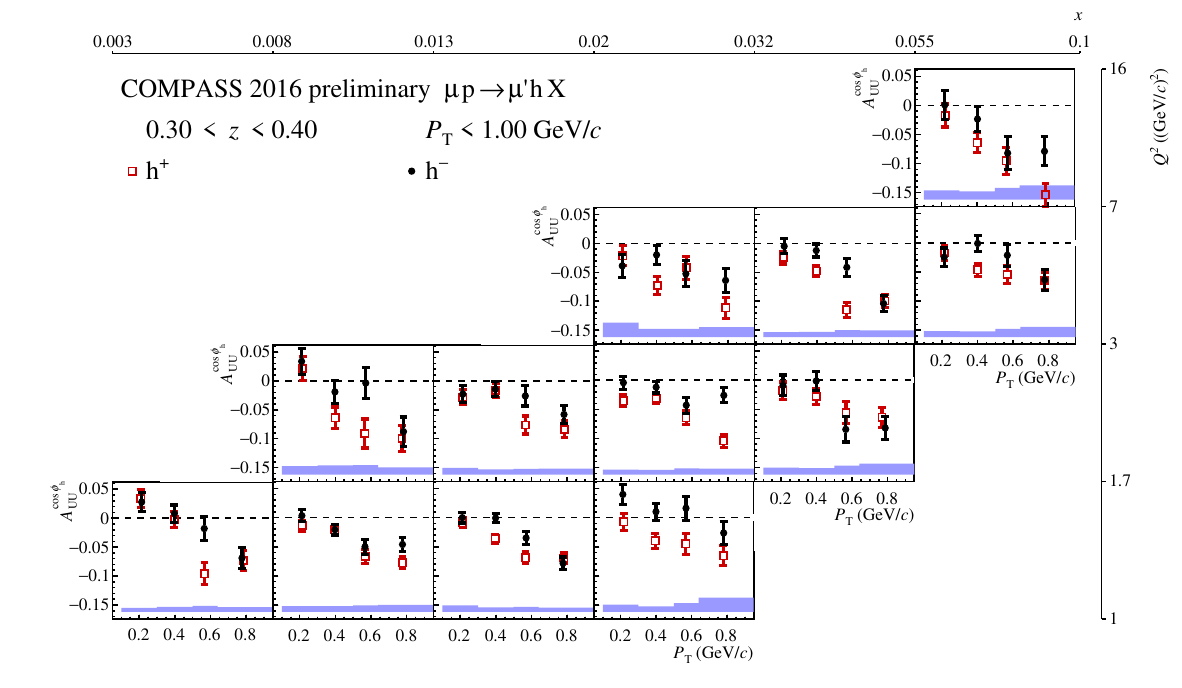}\\
    \includegraphics[width=.84\textwidth]{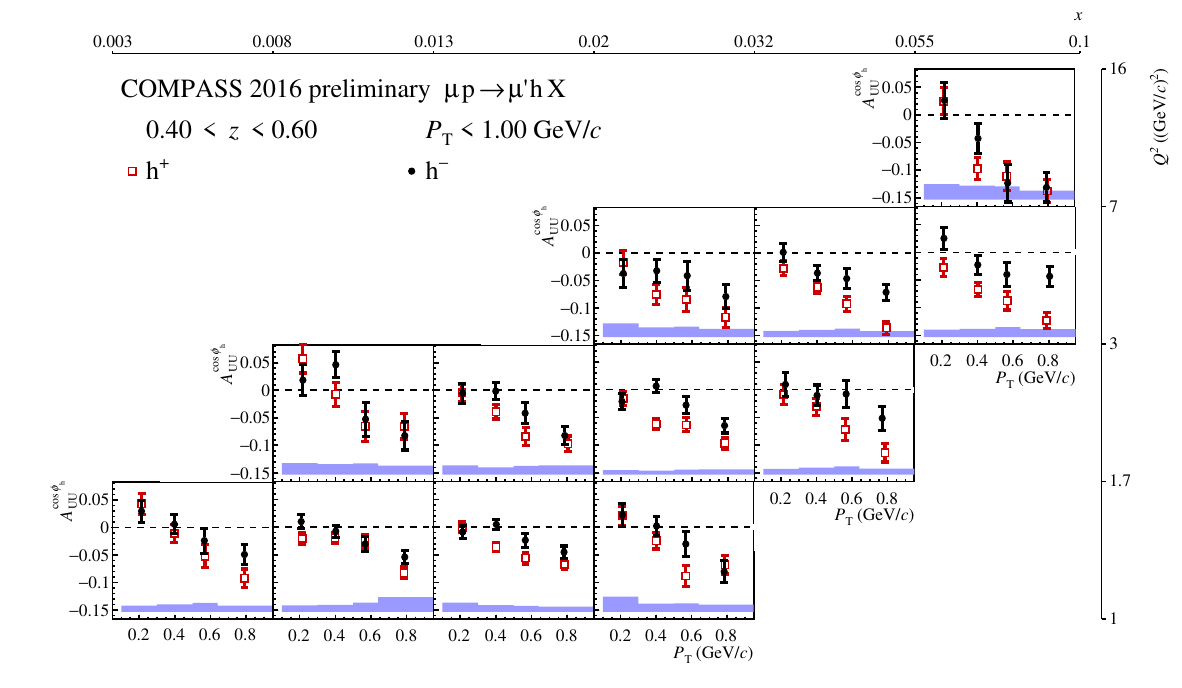}
    \captionof{figure}{Results for $\acos$ in the 4D binning for $0.2<z<0.3$ (top), $0.3<z<0.4$ (middle) and $0.4<z<0.6$ (bottom).}
    \label{fig:4dres_cos}
\end{minipage}%
\hfill
\begin{minipage}{.85\textwidth}
    \centering
    \includegraphics[width=.494\textwidth]{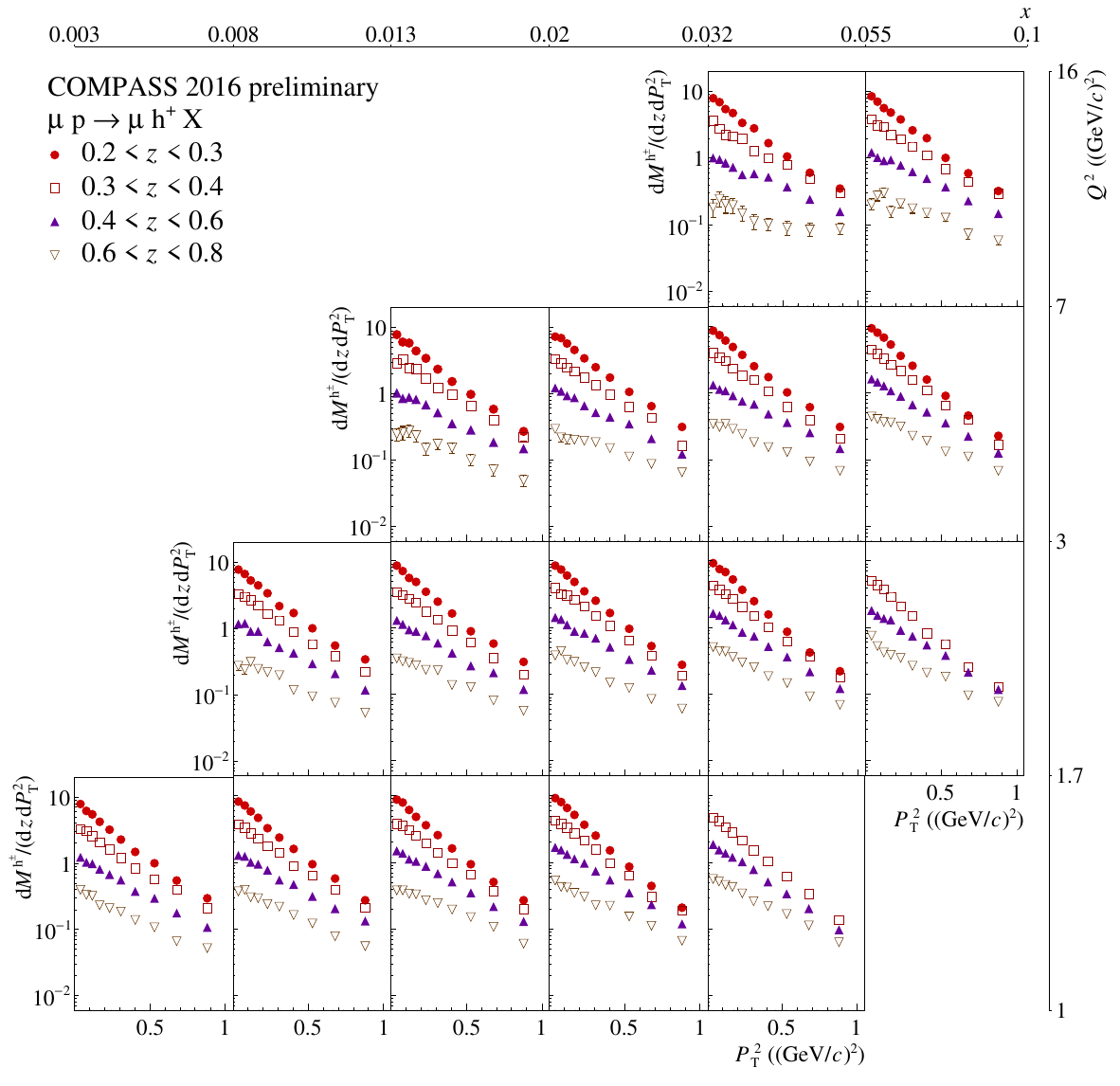}
    \includegraphics[width=.494\textwidth]{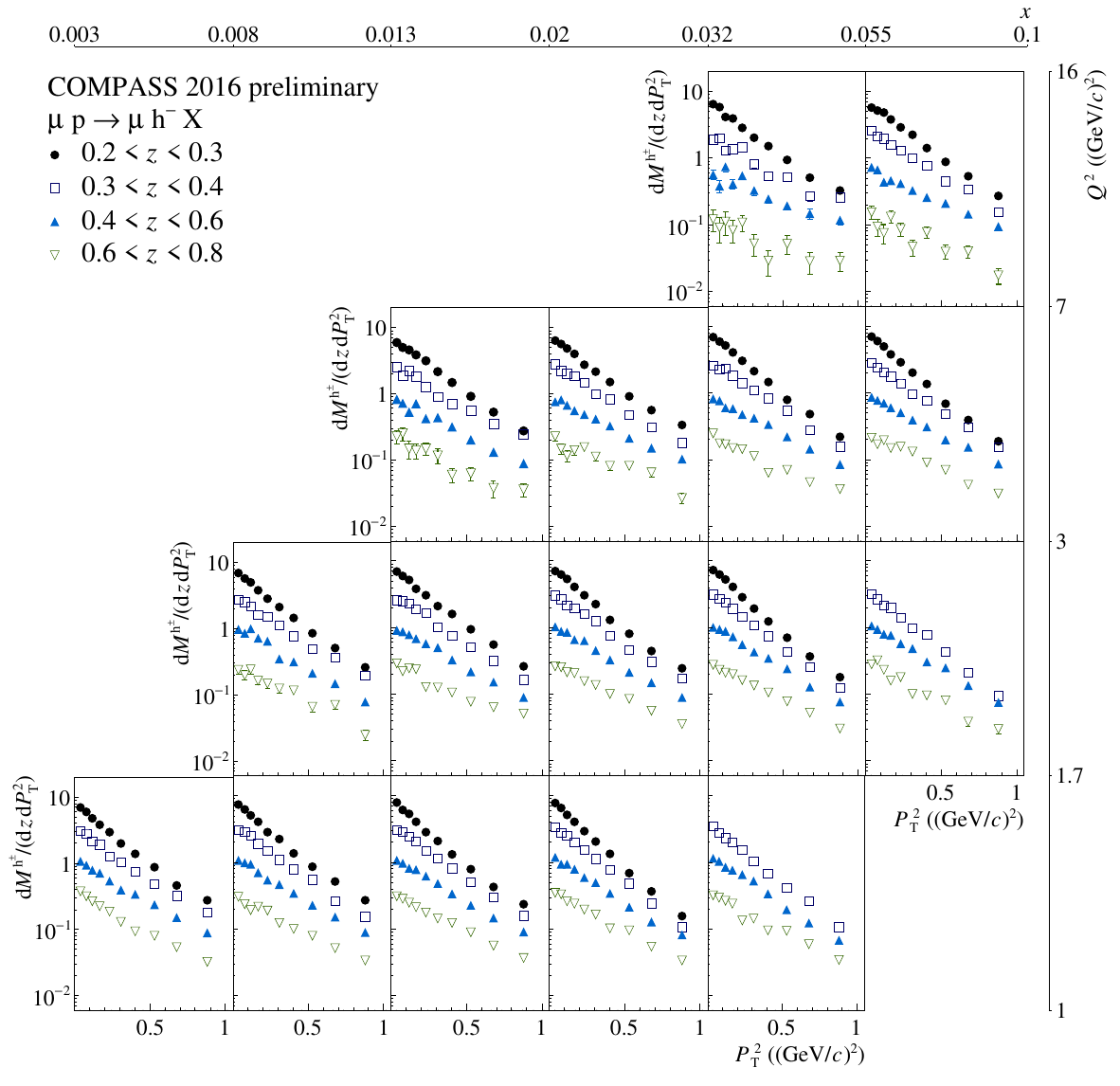}
    \captionof{figure}{Results for $\PhT^2$-multiplicities in the 4D binning.}\label{fig:4dmult}
    \vspace{30pt}
    \includegraphics[width=.494\textwidth]{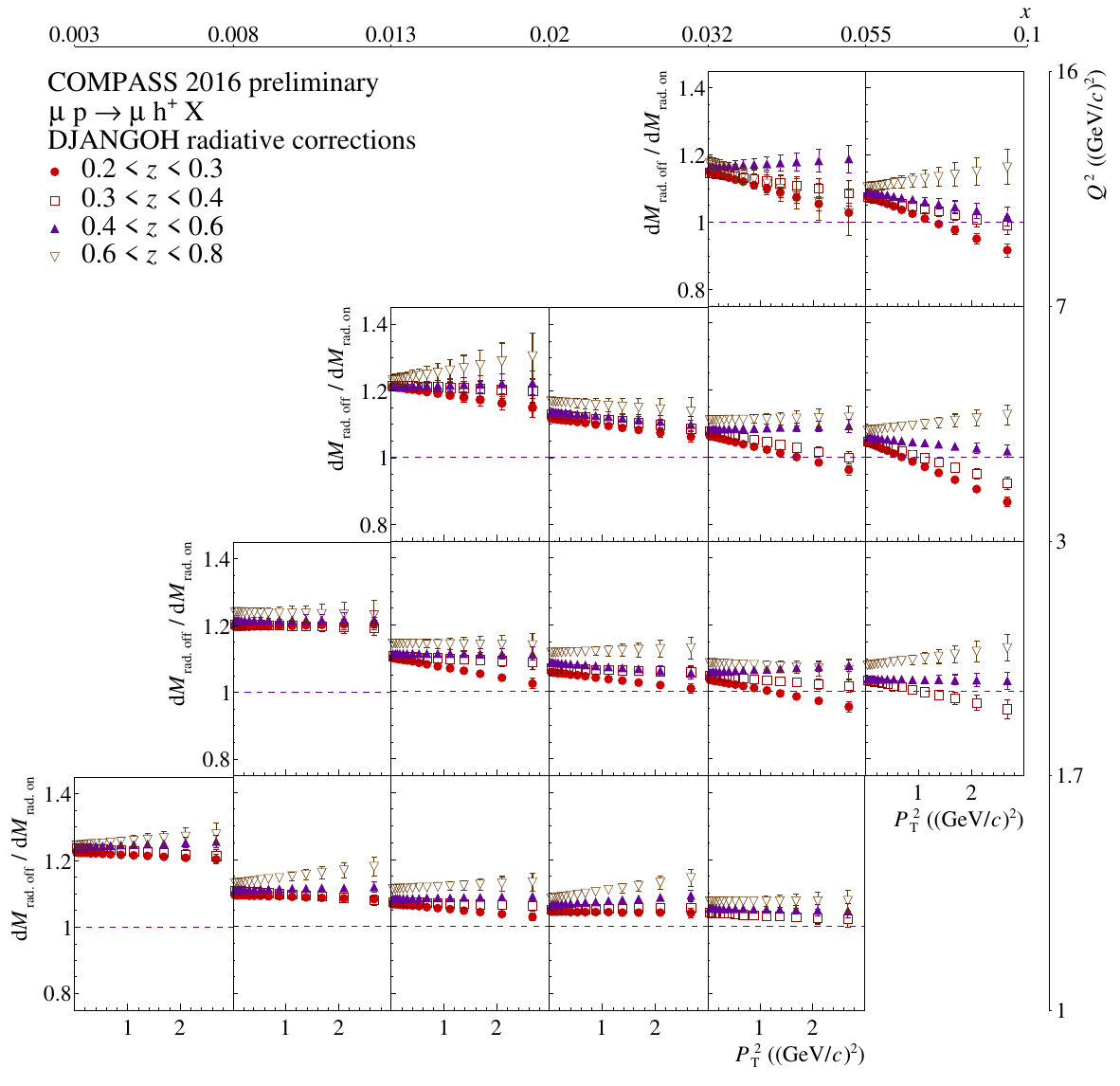}
    \includegraphics[width=.494\textwidth]{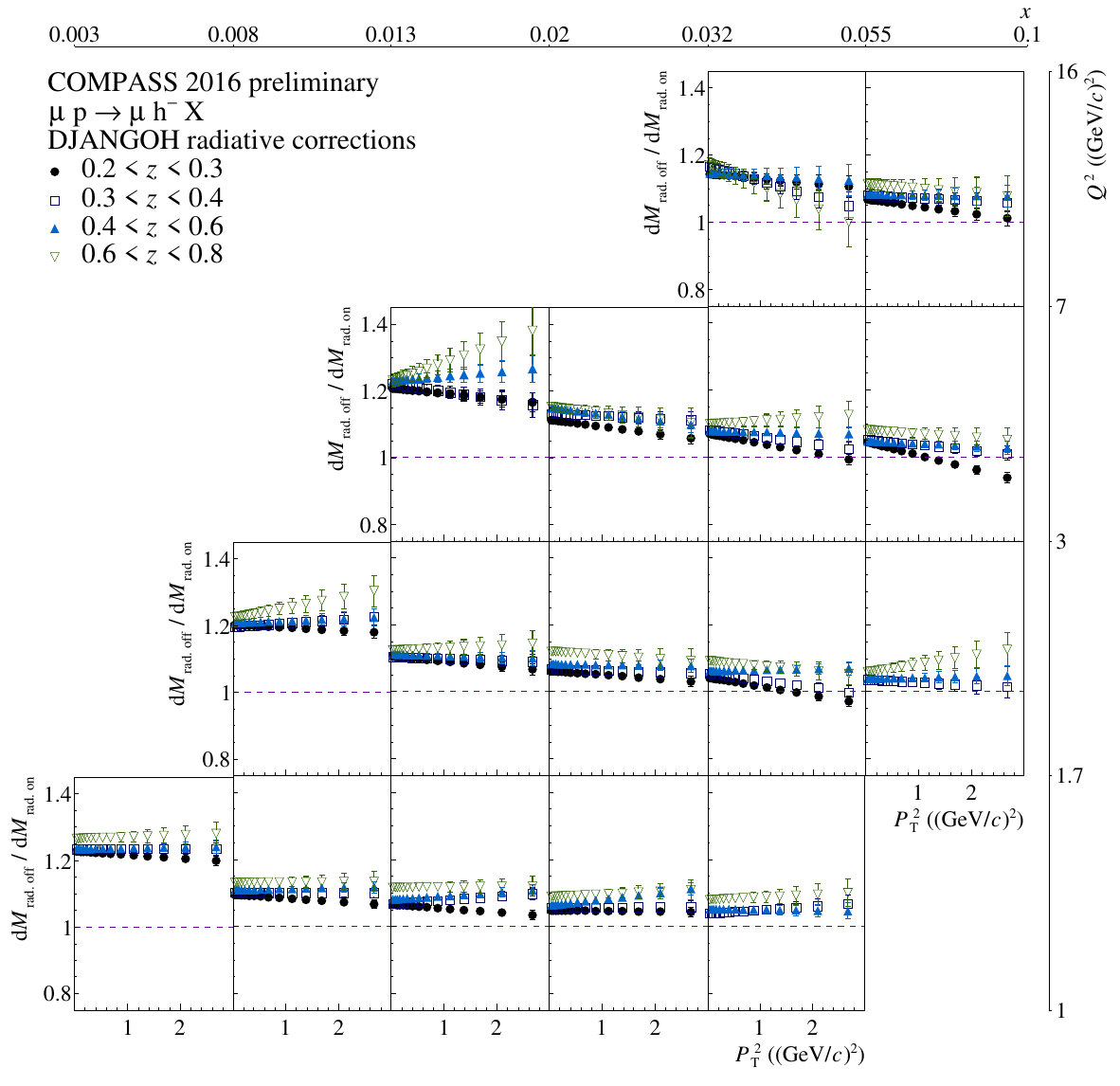}
    \captionof{figure}{Ratio of RC-corrected and uncorrected multiplicities.}\label{fig:4dRC}
\end{minipage}
\end{landscape}

\bibliographystyle{utphys}
\bibliography{bibliography}

\end{document}